\documentstyle[12pt,graphicx]{mrsproc}

\begin{document}

\begin{center}
{\bf Equilibrium structure of decagonal AlNiCo}
\end{center}

\raggedright
\vspace{0.2cm}

\noindent S. Naidu, M. Mihalkovi\v{c}$^1$ and M. Widom,\\
Department of Physics, Carnegie Mellon University, Pittsburgh, PA  15213\\
$^1$also at: Institute of Physics, Slovak Academy of Sciences, 84228 Bratislava, Slovakia

\parindent0.3in

\begin{abstract}
We investigate the high temperature decagonal quasicrystalline phase
of Al$_{72}$Ni$_{20}$Co$_8$ using a quasilattice gas Monte-Carlo
simulation.  To avoid biasing towards a specific model we use an
over-dense site list with a large fraction of free sites, permitting
the simulation to explore an extended region of perpendicular space.
Representing the atomic surface occupancy in a basis of harmonic
functions directly reveals the 5-fold symmetric
component of our data.  Occupancy is examined in
physical and perpendicular space.
\end{abstract}

\section{INTRODUCTION}

AlNiCo exhibits quasicrystalline phases over a range of compositions
and temperatures\cite{Ritsch}.  Of special interest is the Ni-rich
quasicrystalline phase around the composition
Al$_{72}$Ni$_{20}$Co$_8$.  This is a decagonal phase with a period of
4.08~\AA~ along the periodic axis, making it a simple phase relative
to other members of the Al-Ni-Co family. Additionally, it appears to
be most perfect structurally, even though it is stable only at high
temperatures around T=1000-1200K.  Its structure has been extensively
studied experimentally by X-ray diffraction\cite{Takakura,Cervellino}
and electron microscopy\cite{Abe,Pennycook}.  Finally, since
qualitatively accurate pair potentials are available\cite{MGPT},
structural predictions can be made based on total energy
considerations\cite{Mihalkovic}.

An idealized deterministic structure for this phase has been proposed
by studying the total energy\cite{Mihalkovic}.  This prediction
employed a multi-scale simulation method, in which small system sizes
were simulated starting with very limited experimental input, then
rules discovered through the small scale simulations were imposed to
accelerate simulations of larger-scale models.  Although efficient,
this approach leaves open the question of how strongly the final model
was biased by the chosen method.

We adopt a different approach here, starting from slightly different
experimental input and working directly at the relevant high
temperatures.  The experimental input is: (1) the
density~\cite{Takakura}, composition and temperature at which the
phase exists; (2) the hyperspace positions of atomic surfaces (these
are simply the positions for a Penrose tiling, with AS1 at $\nu=\pm 1$
and AS2 at $\nu=\pm 2$); (3) the fact that the quasicrystal is
layered, with space group $10_5/mmc$, and its lattice constants (we
take $a_q=6.427$, $c=4.08$\AA~).  The chief unknowns to be determined
are the sizes, shapes and chemical occupancies of the atomic surfaces.

\begin{figure}
\includegraphics*[width=6in]{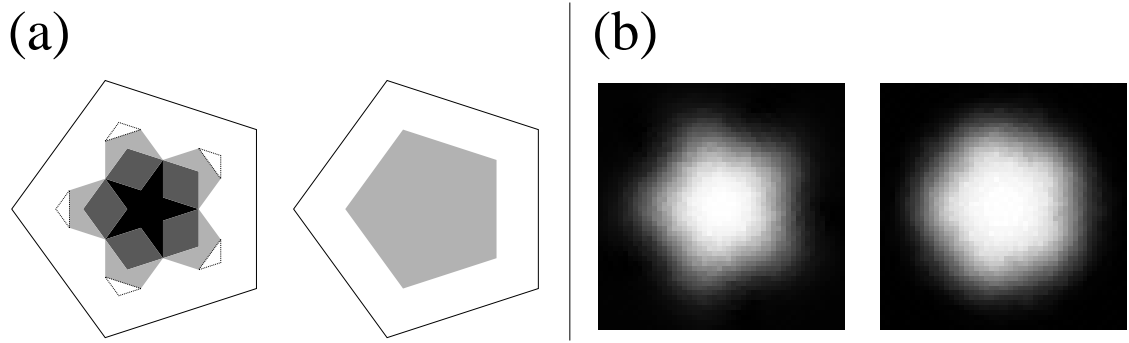}
\caption{\label{fig:model-perp}(a) Oversized atomic surface used in
the present simulation. AS1 (left) at $\nu=\pm1$, AS2 (right) at
$\nu=\pm2$. For comparison the ideal model proposed in
Ref.~\cite{Mihalkovic} is inscribed inside. Color scheme: black=Co,
gray=Ni, light-gray=Al, white=partial Al occupancy.  (b) Total
occupancy of AS1 (left) and AS2 (right) at T=1160K.}
\end{figure}

Like the prior study~\cite{Mihalkovic}, we employ Monte Carlo
simulation using the same electronic-structure derived pair
potentials\cite{MGPT}.  Also, like the prior study, we employ a
discrete list of allowed atomic positions.  However, instead of using
sparse decorations of fundamental tiles, where the configurational
freedom arose largely from flipping of the decorated tiles, in the
present study we employ a fixed site list based on a rich decoration
of fixed tiles.  For a given tiling the density of allowed sites in
our new simulation is much greater than the actual atomic density. The
resulting atomic surface (see Fig.~\ref{fig:model-perp}a) corresponds
to that of a Penrose tiling with tile edge length
$a_q/\tau^3=1.52$~\AA~ (plus a few additional sites such as inside some
fat rhombi).  These atomic surfaces include within them the atomic
surfaces previously proposed on the basis of total energy
calculations~\cite{Mihalkovic}, and analysis of experimental
data~\cite{Takakura,Cervellino}.  This larger atomic surface avoids
any bias towards particular atomic surface shapes.

The following section describes the site list and other
simulation details more precisely.  Then we present site occupancy in real
space and on the atomic surfaces.  We represent atomic surface occupancies
using Fourier-Bessel series, allowing us
to isolate the five-fold symmetric component and filter sampling
noise.  Such a representation could prove fruitful in
crystallographic structure determination as well.  Refs.~\cite{Perez-Mato,JQ}
pursued related goals.

We compare our results with a recent experimental study. Briefly, we
find broad agreement with the sizes, shapes and chemical occupancies
of the atomic surfaces.  A few specific points of disagreement can be
addressed by relaxing our fixed site positions.  Finally, we conclude
with a summary of our results and outlook for future work.

\section{SIMULATION METHOD}

We should perform simulations in the continuum, with no predefined
site list to bias our results. Unfortunately, strong binding of Al to
transition metal atoms (Co and Ni) leads to phase separation into a
CsCl-type crystal plus regions of excess pure Al.  This reflects a
deficiency of our interatomic potentials, which should not be applied
at a transition-metal concentration as high as the 50\% found in
CsCl-type crystals.  Restricting atomic positions a set that does
not include the CsCl crystal inhibits phase separation.

\begin{figure}
\includegraphics*[width = 2 in]{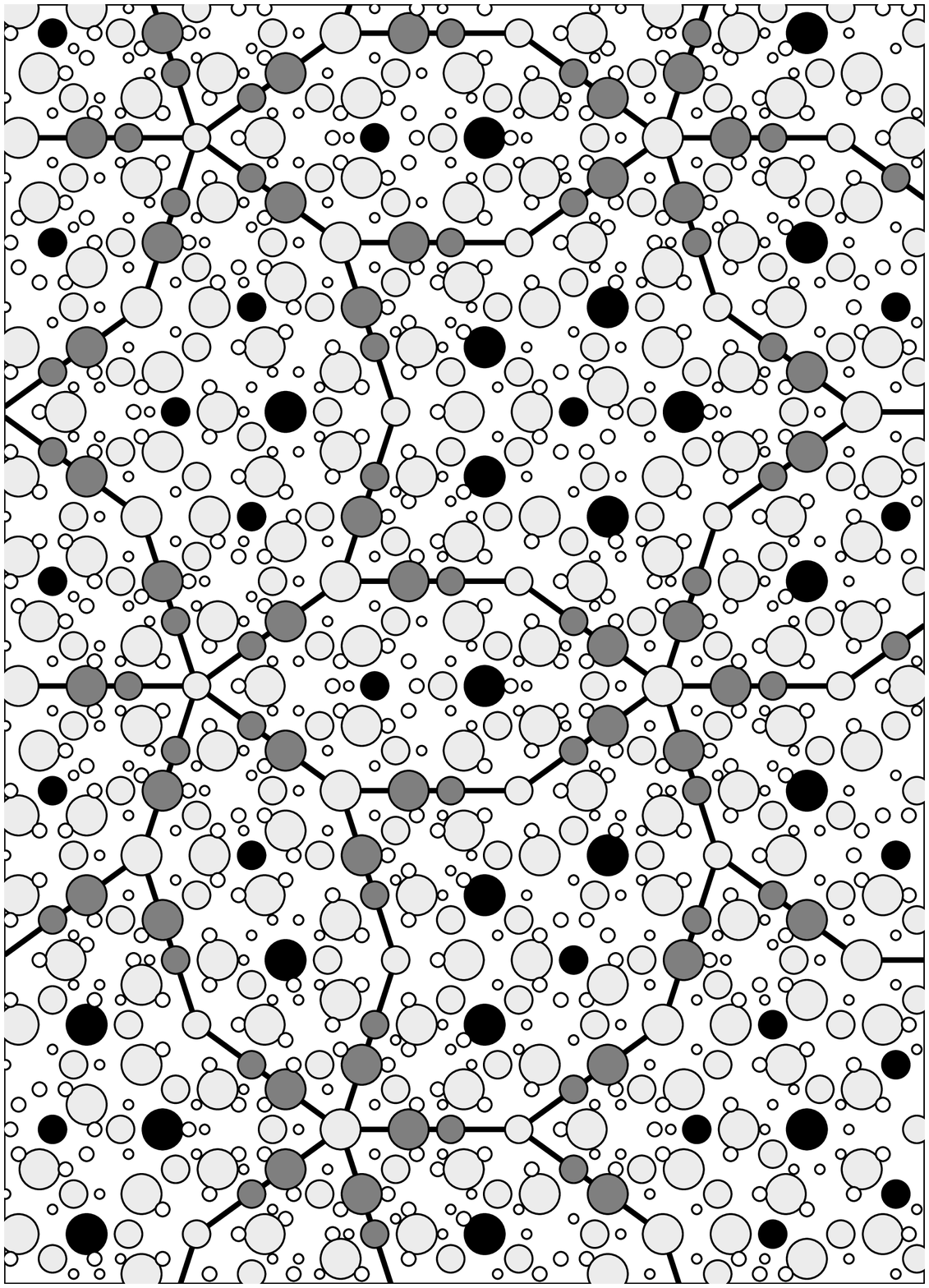}
\includegraphics*[width = 2 in]{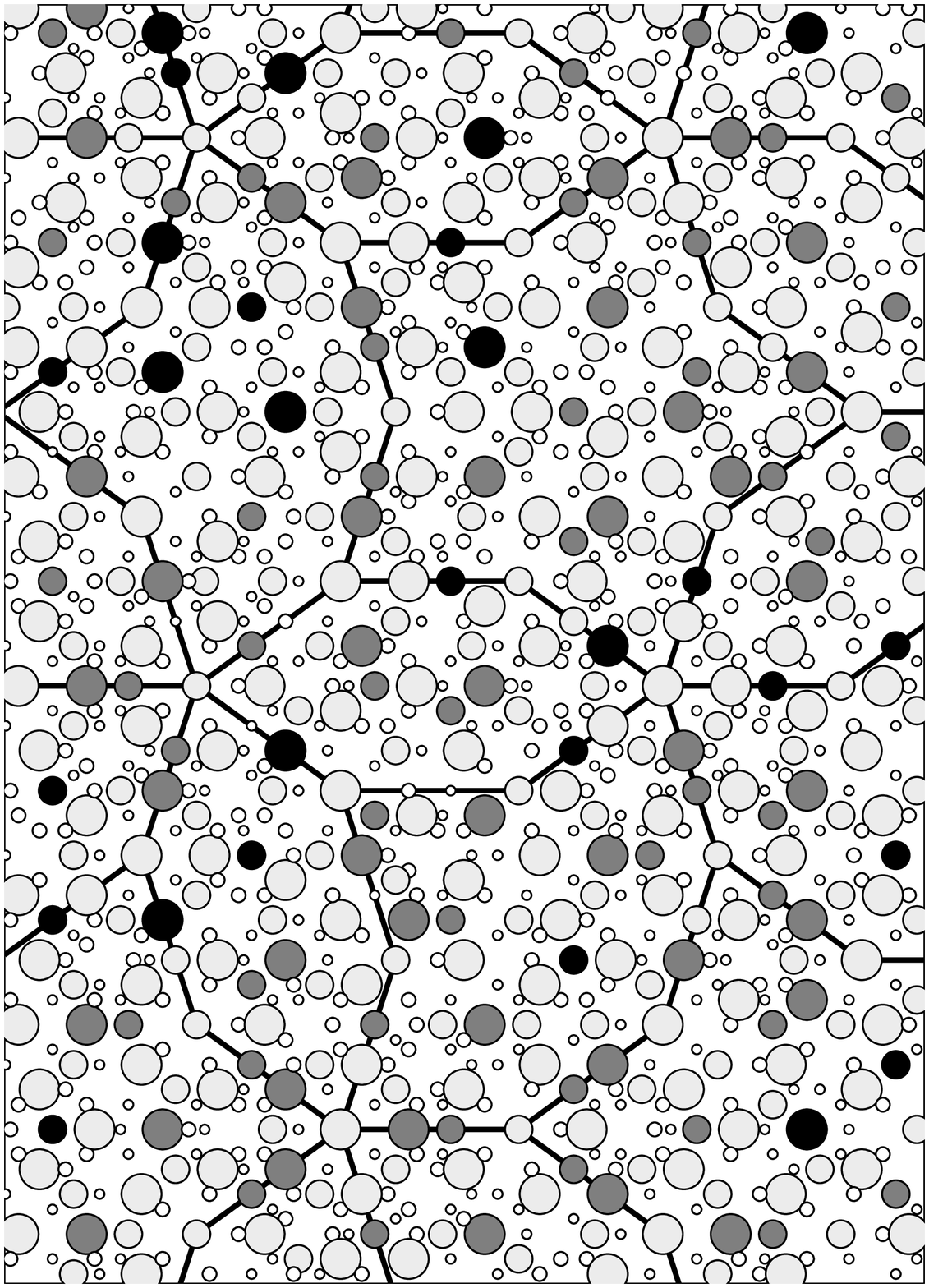}
\includegraphics*[width = 2 in]{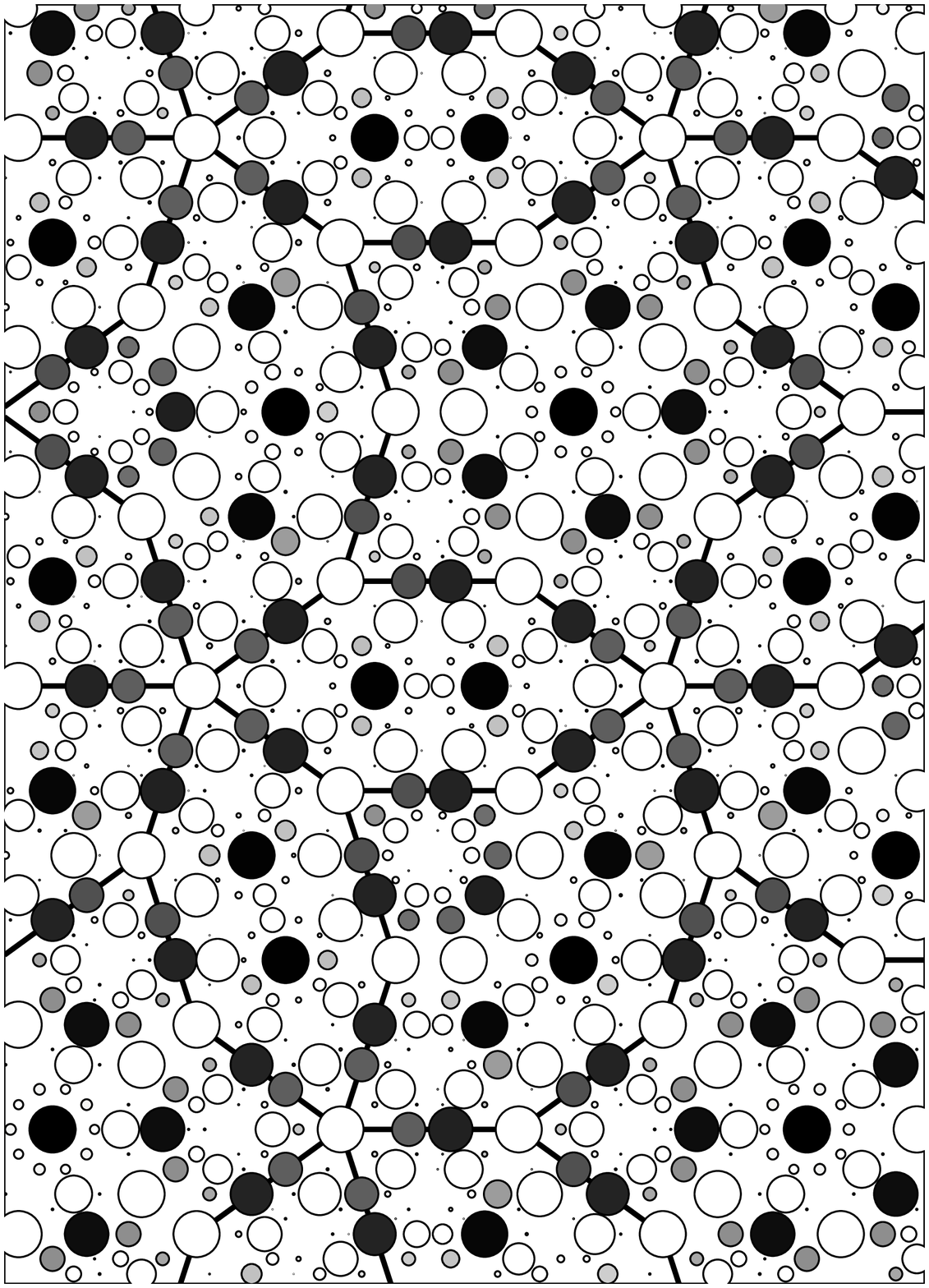}
\caption{\label{fig:model-para}This figure illustrates our fixed site
list in parallel space. Ideal tile decoration (left) and typical
T=1160K configuration (center).  Color scheme as in
Fig~\ref{fig:model-perp}a.  Large/small atoms indicate upper/lower
layer. Very small white circles indicate vacancies. Occupancy data
(right) at T=1160K with atomic size indicating total occupancy and Al
fraction indicated by gray scale (black=100\% TM).}
\end{figure}

Rather than specify these sites in a complex manner whose details may
fail to allow some especially favorable structure, we take a
relatively neutral assignment that still allows quasicrystalline
structures to form.  Specifically we restrict the possible atomic
positions to the vertices of a Penrose lattice whose edge length is
small compared to the spacing between atoms.  The possible atomic
sites fill space with a density 2.55 times greater than the actual
atomic density, so the majority of sites are empty in any particular
configuration.  Our prior study of prefered structures proposed an
idealized arrangement of atoms on an HBS (hexagon-boat-star) tiling of
edge length $a_q$, as illustrated in Fig.~\ref{fig:model-para}.

Our Monte Carlo simulation distributes atoms among these sites in a
manner consistent with thermal equilibrium at a given temperature.  We
focus on the temperature T=1160K, which lies within the stability
range for the Ni-rich decagonal phase~\cite{Ritsch}.  We simulate a
low phason strain quasicrystal approximant, with lattice parameters
$159.38\times 83.79\times 4.08$~\AA$^3$, containing 9776 lattice sites
and holding 3838 atoms.  We initially decorated the structure randomly
at the desired density and composition.  We then lowered the
temperature in steps from an initial T=4000K, annealing at each
temperature, until we reached our data collection temperature of
T=1160K.

Our data analysis differs from conventional approaches to quasicrystal
atomic surface modeling.  It is common to model atomic surface
occupancy by breaking the atomic surface up into polygonal domains
piecewise constant occupancy in each domain. This approach highlights
the manner in which occupancy depends on the local real space
environment, which changes discontinuously as a function in
perpendicular space. Unfortunately such a piecewise continuous
representation does not properly capture the effects of chemical
disorder and phason fluctuations characteristic of a quasicrystal at
high temperatures.

We show below that the mean occupancy statistics actually vary quite
smoothly over the atomic surface.  To represent smoothly varying
atomic surface occupancy functions it is helpful to introduce a
complete basis set of analytic functions and expand the occupancy in
this basis.  Owing to the axial symmetry of the atomic surfaces it is
handy to introduce polar coordinates $(r_{\perp},\theta)$.  The
natural function basis set for polar coordinates are combinations of
cylindrical Bessel functions multiplied by complex exponentials.
Thus, on a given atomic surface we express the occupancy by chemical
species $\alpha$ as
\begin{equation}
\rho_{\alpha}(r_{\perp},\theta) =
\sum_{m,n} A^{\alpha}_{m,n} J_m(k_{m,n} r_{\perp}) e^{i m \theta}.
\label{eq:bessel}
\end{equation}
The $m$ index represents the angular mode frequency.  For a given
angular frequency $m$, the sum over $n$ allows the representation of
an arbitrary radial variation, with radial frequencies $k_{m,n}$ such
that all Bessel functions share a common zero outside our atomic
surface.  To analyze our data we invert eq.~(\ref{eq:bessel}) to
obtain the coefficients $A^{\alpha}_{m,n}$ using orthonormality of the
basis set.  We then filter out the components whose angular variation
lacks 5-fold symmetry by restricting $m$ to multiples of 5.

\section{RESULTS}

Each Monte Carlo step consisted of many attempted jumps (of up to
$6.5$~\AA~ in length) per atom.  An attempted jump to an occupied site
means an attempt to swap the two atoms.  We recorded the detailed
atomic site occupancy for 9000 configurations separated by
sufficiently many Monte Carlo steps that the configurations become
uncorrelated.

Phason fluctuations such as bowtie flips~\cite{Mihalkovic}
cause the center of gravity of the
structure to drift in perpendicular space.  Note that the lattice
sites do not move during a phason flip, only their atomic occupancy
changes.  A bowtie flip in our ideal model interchanges
two NiNi pairs with two AlCo pairs.  To avoid smearing caused by this drift,
which could obscure small-scale atomic surface details,
we recentered the perp space center of gravity for each configuration
before projecting onto the Fourier-Bessel coefficients.  Without
recentering the atomic surface occupancy
functions would be weaker and broader.

\begin{figure}
\hspace{1.5in}
\includegraphics*[width = 1.18 in, angle=90]{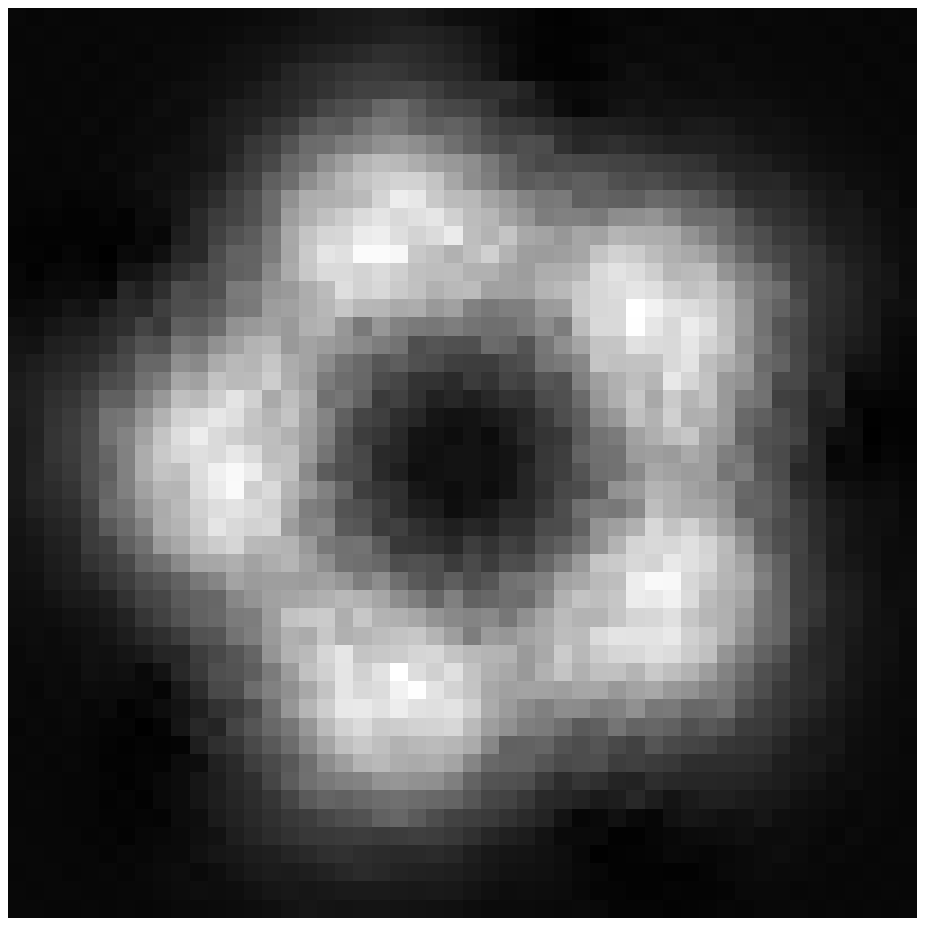}
\includegraphics*[width = 1.18 in, angle=90]{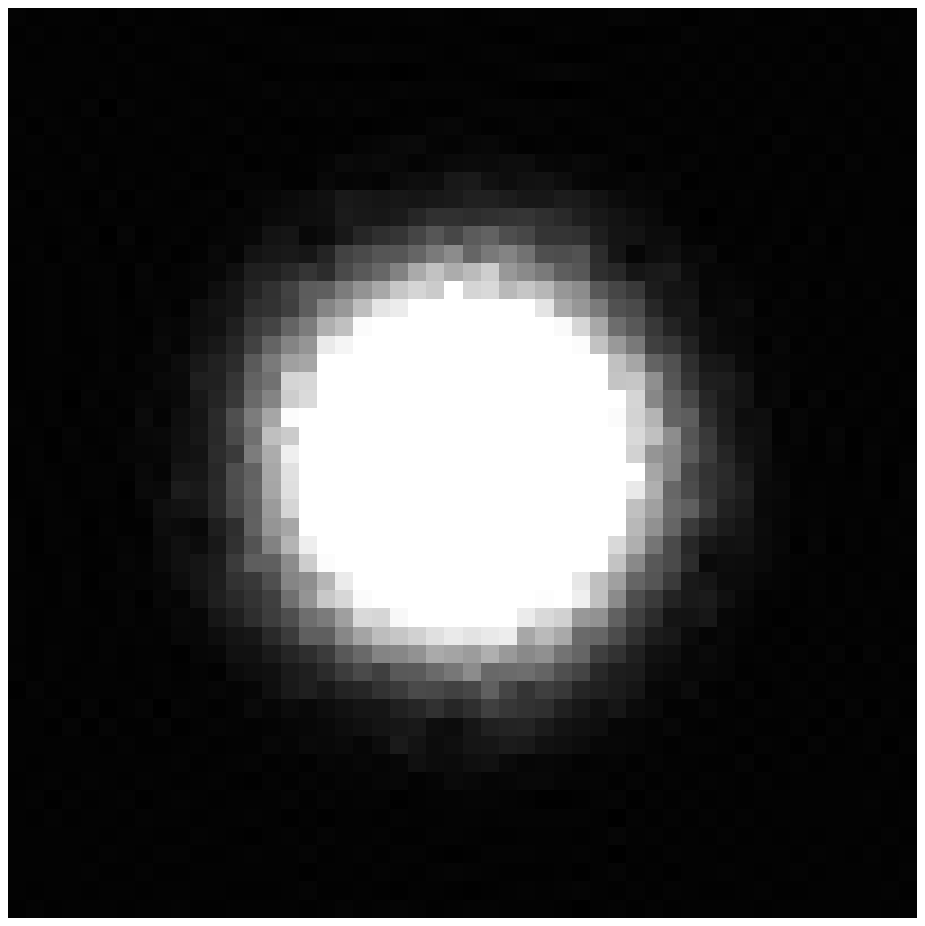}
\includegraphics*[width = 1.18 in, angle=90]{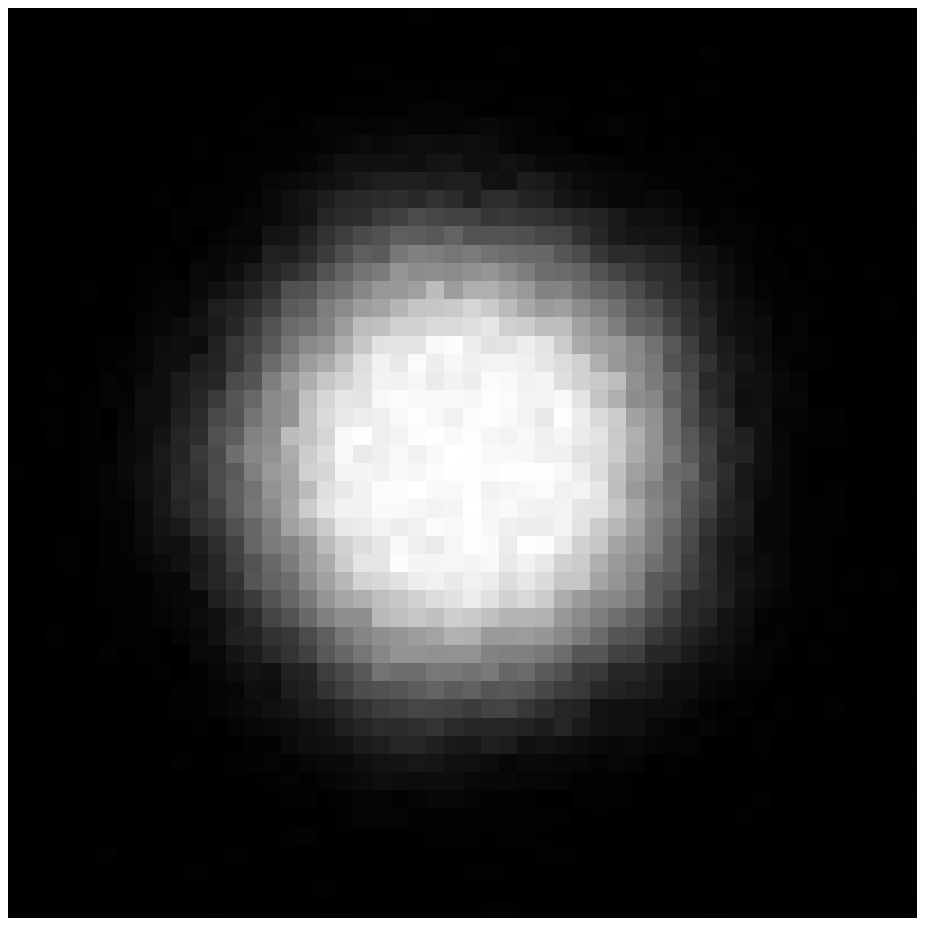}
\caption{\label{fig:AS1AlNiCo}This figure illustrates the chemical
occupancy of the $\nu=\pm 1$ atomic surfaces for Al, Ni and Co atoms
(left to right). The gray scales are proportional to occupancy, scaled
to the maximum for each species.}
\end{figure}

As in our prior~\cite{Mihalkovic} low temperature study, we find that
the HBS tiling effectively describes the quasiperiodic layers, but the
prefered chemical decoration of the tiles now becomes highly variable
at certain sites.  Fig.~\ref{fig:model-perp}(b) displays the average
occupancy distribution on each atomic surface.  The AS1 occupancy is a
combination of all three chemical species.  The AS2 occupancy is
virtually 100\% Al.  The breakdown of AS1 occupancy among Al, Ni and
Co is shown in Fig.~\ref{fig:AS1AlNiCo}.  A striking feature of all
these occupancy plots is the smooth, continuous variation of
occupancy.  It is clearly not appropriate to model these densities
using piecewise constant functions with polygonal boundaries, even
though five-fold symmetry is present.  Ni and Co are strongly mixed
with very similar distributions, concentrated at the center of
AS1. Because the amount of Ni present exceeds the amount of Co, the Co
occupancy vanishes more quickly than the Ni occupancy away from the
center, while the Ni occupancy remains more nearly constant. In the
region where the TM occupancy drops off it is largely replaced with Al
so the total occupancy remains close to 100\% in this crossover
region.

In parallel space atoms in the TM to Al crossover region
correspond to tile edges.  Effects of this gradient from TM to
Al occupancy can be seen in Fig.~\ref{fig:model-para} (right), near tiling
vertices at which two or more edges join in a $\pm72^{\circ}$ angle.
The pair of occupied sites on these tile edges show unequal Al/TM
occupancy.  The site closer to the vertex contains relatively less TM
and more Al than the site further away.  The site closer to the vertex
in parallel space lies in the atomic surface region that is further
from the center in perp space.  Note also that bowtie phason
flips that replace TM--TM pairs with Al--TM pairs on tile edges are one
type of excitation that replace the TM site close to a 72$^{\circ}$
vertex with an Al atom.  We understand these effects energetically
because the inner edge site is in a TM-rich environment.  Al--TM bonds
are strongly favorable so this is a likely site for Al occupancy.

Far from the atomic surface centers, where the occupancy falls
smoothly to zero, the atoms are mainly Al.  These highly mobile atoms
are in locations in parallel space where there are multiple ideal
sites, too close to be simultaneously occupied, but each one with a
rather similar local environment so that there is no strong energetic
preference among the sites, for example the pair of symmetry-related
sites near the center of the hexagon tile.  Other examples are rings
of 10 sites surrounding the ideal Co position at the centers of boat
and star tiles.  These rings can hold at most three Al atoms.  Jumps of
Al atoms among these sites correspond to phason tile flips of the
underlying very small 1.52~\AA~ rhombus tiling.

\section{COMPARISON WITH EXPERIMENT}

We compare primarily with the refinement by Takakura~\cite{Takakura}.
We first point areas of general agreement between our results and the
experimental data, then discuss the main points of disagreement.
Our simulated occupation probabilities generally agree well
with experiment.  The approximate sizes and shapes of our atomic
surfaces agree with Takakura.  We agree that AS1 contains transition
metal atoms within a central region (Takakura orbit numbers 1-5),
separated by a fully occupied region of mixed Al/TM (orbits 6 and 7)
and finally radial spokes in which Al occupancy diminishes from full
to partial (orbits 8-10).  We agree that AS2 (orbit numbers 11-23)
contains primarily Al atoms.  Takakura finds full occupancy of orbit
numbers 14 and 16, but only 82\% occupancy of orbit numbers 15 and 17,
which are locally fairly similar (these are the Al sites adjacent to
tile edge TM atoms).  Our results qualitatively support this, with
occupancy about 90\% corresponding to orbit numbers 14 and 16,
dropping to around 70-80\% for orbit numbers 15 and 17.

Now consider discrepancies between our results and
experiment. Orbits 19 and 23, localized around special points
on the fringes of AS2 are fully occupied in the refinement, while we
find partial occupancy.  In parallel space these correspond to
pairs of ideal sites midway between TM atoms.  Partial occupancy is
forced because the spatial separation of these ideal points
(2.25~\AA) is too small to allow both to be simultaneously occupied.
However, if we consider multilayer strucures, with $c=4\times
4.08$~\AA, and allow structural relaxation, we know~\cite{HMW} it
is possible to occupy 3 out of every 4 such sites per 8.16~\AA~.
The lack of structural relaxation thus causes an error in our
occupancy for this orbit.

Takakura's fully occupied orbits numbers 20-22 are located on the
corners of AS2.  We find 60-80\% occupancy instead of full occupancy,
consistently with our smoothly decreasing occupation probability.
Takakura also finds a small fraction (about 20\%) of TM atoms on
orbits 20 and 21, while we find only Al.  There is no strong energetic
reason that TM atoms should not occupy those sites.  Perhaps the TM
occupancy of orbit number 20 is related to the large atomic
displacement of this orbit Takakura found during the refinement.

Surprisingly, Takakura finds 50\% occupancy on orbit number 11, at the
very center of AS2. In parallel space this site corresponds to HBS
tile vertices at which five edges meet.  We find it is fully occupied,
similar to all other HBS tile vertices.

\section{CONCLUSION}

We presented a method to compute atomic surface occupancy statistics,
and applied it to the basic Ni-rich decagonal phase of AlNiCo.  This
work launches a promising set of possible future research directions.
The method can be applied to other decagonal phases and also to
icosahedral phases.  The main limitation is the availability of
interatomic potentials.  Further, a great deal remains to be done even
within AlNiCo, as we now discuss.

Our simulations generate a complete set of equilibrium configurations,
in which the precise location of every atom is known.  The analysis
above only presents the atomic surface occupation probabilities, but
does not address correlations among the partially occupied sites.
Such correlations are contained in the Patterson function, which is
also known experimentally~\cite{Steurer} and can be computed from our
existing data.  Actual atomic configurations contain more information than the
atomic surface occupancies, because all correlations are reflected in
the actual configurations.  Fig.~\ref{fig:model-para} (center)
illustrates an atomic arrangement from our
simulation.  Comparing with the ideal model (left) reveals a great
deal of disorder, including phason flips, chemical disorder and
vacancies.

Among the prominent correlations are strong anticorrelation of
too-close sites.  The deep clefts between arms of AS1
correspond to sites that are close in parallel space to sites
near the corners of AS2.  When these AS2 sites are occupied sites
within the clefts of AS1 are empty.  When these AS2 sites are empty
then we find partial occupancy inside the clefts of AS1.
Another anticorrelation occurs because the pairs of ideal
sites midway between pairs of TM atoms (e.g. Takakura orbit number 19
and 23) are too close for simultaneous occupancy.  This
anticorrelation arises because we used only ``ideal'' positions
projected into parallel space from flat atomic surfaces.  In reality
the atomic positions are displaced somewhat away from the ideal sites,
and we showed above that such displacements can have a substantial
influence on occupancy statistics in cases where the ideal sites are
just slightly too close together.  Indeed, relaxations of our ideal
structures, using either pair potentials~\cite{HMW} or full
{\em ab-initio} calculations result in displacements similar
to those reported experimentally~\cite{Takakura}.

Finally, our simulation used only a single 4.08~\AA~ repeat unit along
the 10-fold symmetry axis.  Owing to our periodic boundary conditions,
swapping a pair of atoms in the single repeat implies a simultaneous
swap of all pairs of atoms in infinite columns extending above and
below the original pair in our single repeat.  But when we calculate
the energy for the swap, we only include the energies of the original
atoms.  This type of finite-size error needs to be eliminated, as it
miscalculates the energies of phason flips.

Simply stacking two repeats above each other to make an 8.16~\AA~
unit, however, usually overestimates the energy cost of atom swaps
corresponding to bowtie phason flips.  This is because uncorrelated
flips between repeat units create phason stacking faults whose energy
cost is large in the absence of atomic relaxation but is small, and
can even be negative, when relaxation is included.  We must include
relaxed atomic positions to achieve a realistic model of
three-dimensional phason disorder.

\section{ACKNOWLEDGEMENTS}

We acknowledge discussions with C.L. Henley support by NSF grant DMR-0111198.

\end{document}